\documentclass[dvips]{article}

\usepackage{amssymb}
\usepackage{icrctc07}

\newcommand{\mc}{\mathcal}
\newcommand{\mn}{\mathnormal}

\newcommand{\ck}{Cerenkov~}

\title{The AMS-RICH velocity and charge reconstruction}
\shorttitle{RICH reconstrution}

\authors{%
M. Aguilar-Benitez$^{1}$,
L. Arruda$^{2}$,
{\bf F. Barao$^{2}$},
B. Baret$^{3}$, 
A. Barrau$^{3}$, 
G. Barreira$^{2}$, 
E. Belmont$^{4}$, 
J. Berdugo$^{1}$,
J. Borges$^{2}$,
M. Buenerd$^{3}$, 
D. Casadei$^{5}$,
J. Casaus$^{1}$,
E. Cortina$^{1}$,
M. Costado$^{7}$,
D. Crespo$^{1}$,
C. Delgado$^{7}$,
C. Diaz$^{1}$,
L. Derome$^{3}$,
P. Gon\c{c}alves$^{2}$,
R. Garcia-Lopez$^{7}$,
C. de la Guia$^{1}$,
A. Herrero$^{7}$,
E. Lanciotti$^{1}$,
G. Laurenti$^{5}$,
A. Malinin$^{6}$,
C. Mana$^{1}$,
J. Marin$^{1}$,
M. Mangin-Brinet$^{3}$,
G. Martinez$^{1}$,
A. Menchaca-Rocha$^{4}$,
C. Palomares$^{1}$,
R. Pereira$^{2}$,
M. Pimenta$^{2}$,
A. Putze$^{3}$,
Y. Sallaz-Damaz$^{3}$,
E.S. Seo$^{6}$,
I. Sevilla$^{1}$,
A. Torrento$^{1}$,
M. Vargas-Trevino$^{3}$,
O. Veziant$^{3}$ 
}
\shortauthors{F. Barao et al.}
\afiliations{%
$^1${CIEMAT}, Avenida Complutense 22, E-28040 Madrid, Spain\\
$^2${LIP}, Avenida Elias Garcia 14-1, P-1000 Lisboa, Portugal\\
$^3${LPSC}, IN2P3/CNRS, 53 av. des Martyrs, 38026 Grenoble cedex, France\\
$^4$Instituto de Fisica, UNAM, AP 20-364 Mexico DF, Mexico\\
$^5$University of Bologna and INFN, Via Irnerio 46, I-40126 Bologna, Italy\\
$^6$University of Maryland, College Park, MD 20742, USA\\
$^7${IAC}, C/Via Lactea s/n E28295 Tenerife, Spain
}
\email{barao@lip.pt}

\abstract{%
The AMS detector, to be installed on the International Space Station,
includes a Ring Imaging Cerenkov detector with two different radiators,
silica aerogel (n=1.05) and sodium fluoride (n=1.334). This detector is
designed to provide very precise measurements of velocity and electric
charge in a wide range of cosmic nuclei energies and atomic numbers.
The detector geometry, in particular the presence of a reflector for
acceptance purposes, leads to complex Cerenkov patterns detected in a
pixelized photomultiplier matrix. The results of different reconstruction
methods applied to test beam data as well as to simulated samples are
presented.
To ensure nominal performances throughout the flight, several detector
parameters have to be carefully monitored. The algorithms developed to
fulfill these requirements are presented.
The velocity and charge measurements provided by the RICH detector
endow the AMS spectrometer with precise particle identification
capabilities in a wide energy range. The expected performances on light
isotope separation are discussed.
}

\begin{document}
\maketitle
\vspace{-0.2cm}
\section{The AMS and  RICH detectors}
\vspace{-0.3cm}
The Alpha Magnetic Spectrometer
\cite{bib:ams02-icrc05} 
(AMS) is a high energy physics experiment that 
will be installed on the International Space Station (ISS) by the year 2009, where 
it will operate for a period of at least three years. 
It is a large acceptance (\textrm{$\sim 0.5$ m$^2$sr}) superconducting magnetic spectrometer
able to detect in a wide kinematic range (from a few hundred MeV up to TeV region) singly charged 
particles, charged nuclei 
and $\gamma$ rays.
The long time exposure in space will allow AMS to collect an unprecedented large
data sample and to extend by orders of magnitude the sensitivity reached by previous 
experiments on dark matter and antimatter searches.
In addition, the measurement of the cosmic-ray abundances up to the TV region and in a wide 
charge range (up to around iron) will contribute to a better description of cosmic ray production,
acceleration and propagation mechanisms, essential for a full understanding of the background spectra 
on dark matter searches.
Information about the density of the interstellar medium traversed by the 
cosmic rays and their confinement time can be derived from the isotopic composition of 
secondary cosmic rays, produced by fragmentation during the cosmic ray transport in the galaxy.
For instance, the relative abundances of deuterium and helium-3 isotopes reflect the transport 
history along the galaxy of protons and heliums, while the beryllium-10 radionuclide  accounts for 
the time confinement.      
Current measurements are performed at relatively low energies (T $\lesssim 1$ GeV/n) and based on 
small statistics. 

Particle identification with AMS-02 relies on a very precise determination 
of the magnetic rigidity, energy, velocity and electric charge.
In the AMS spectrometer, the momentum is obtained from the information  
provided by the silicon tracker
with a relative accuracy of $\sim$2\% up to $10$ GeV/c/n. 
Isotopic mass separation over a wide range of energies requires,
in addition to an accurate momentum measurement, 
a velocity determination with low relative uncertainty 
as 
\mbox{$\Delta m/m = ( \Delta p / p ) \oplus \gamma^2 ( \Delta \beta / \beta )$}.

For this purpose, the AMS spectrometer includes a Ring Imaging Cerenkov detector 
(RICH)~\cite{bib:rich-icrc}
operating between the time-of-flight and electromagnetic calorimeter detectors.
It was designed to provide measurements of the velocity for singly charged particles with a 
relative uncertainty of 0.1\% and of the nuclei electric charge up to iron.
Moreover, it will provide AMS with an additional contribution to the electron/proton separation.  
For the isotopic separation, the RICH detector will cover a kinetic energy region ranging from 
0.5 GeV/n up to around 10 GeV/n for A $\lesssim 10$. 
The RICH has a truncated conical shape with a top radius of 60 cm, a bottom radius of 67 cm, 
and a total an expansion height of 47 cm. 
It covers 80\% of the AMS magnet acceptance.
It is a proximity focusing device with a dual solid radiator configuration on the top made of 
25 mm thick aerogel and 5 mm thick sodium fluoride (NaF) tiles, the later being crossed 
by  $\sim 11$\% of the events.
The photon detection is made with an array of multianode (16) shielded Hamamatsu tubes (R7600-00-M16), 
optically coupled to light guide acrylic pipes.

\vspace{-0.2cm}
\section{Velocity and charge reconstruction}
\vspace{-0.3cm}
Charged particles crossing the radiator material of refractive index $n$ and with a velocity
larger than $1/n$, emit photons.
The aperture angle ($\theta_c$) of the photons with respect to the radiating particle 
direction depends on the particle velocity $\beta$, 
\begin{equation}
\cos\theta_c = \frac{1}{\beta ~n}
\end{equation} 
The detected ring photon pattern reflects the geometry of the photomultiplier detection matrix 
and the interactions suffered by the emitted photons along their 
path to the readout matrix: radiator interactions (Rayleigh scattering, absorption), 
surface optical effects (reflection and refraction) and
light guide efficiency.
Consequently, from the point of view of the expected Cerenkov pattern, a typical event will 
be composed of aligned photons, strongly uncorrelated scattered photons and detector noise.   

Two different approaches were implemented for the Cerenkov ring reconstruction.
One was based on single hit reconstruction 
~\cite{bib:rich-ciematrec}
and the other on a maximum likelihood method
~\cite{bib:rich-liprec}.
In the former method a value of $\beta$ is reconstructed for every detected hit.
The method is purely geometrical and relies in a set of analytical equations that relate the
detection point with the Cerenkov angle ($\theta_c$) and 
the particle coordinates.
The possibility of the photon being reflected is taken into account.  
Next, the most probable cluster of hits is searched and the final velocity is computed as 
a mean of the clusterized hit $\beta$ values, weighted with measured signal amplitude
(photon multiplicity).  

\begin{figure}
\begin{center}
\noindent
\vspace{-0.6cm}
\includegraphics*[width=0.5\textwidth,angle=0,clip]{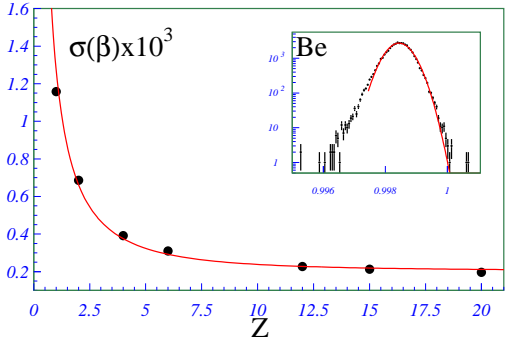}
\vspace{-0.9cm}
\caption{Velocity reconstruction accuracy. 
\label{fig:betarec-vs-z}}
\end{center}
\vspace{-0.8cm}
\end{figure}

In the other reconstruction approach, the algorithm incorporates a probability density
function for the detected hits.
The residuals of the signal hits distribute according to a double gaussian function 
whose widths are directly related to the  
pixel size and granularity, radiator thickness, chromaticity effects and 
aerogel forward scattering.
The existence of the second gaussian, accounting for a natural enlargement of the hit
residuals from forward scattering, makes possible the description of the signal to larger
hit distances and endows a better algorithm efficiency and a lower sensitivity to noisy
hits.  
The signal probability density function $S(r)$ associated to a hit $i$ at a distance $r$ from the 
hypothetical ring ($\theta_c$) is as follows:
\[
S(r) = \alpha_1 \; G_1(\sigma_1) + \alpha_2 \; G_2(\sigma_2)
\]
where $\alpha_1$ and $\alpha_2$ are respectively $\sim 3/4$ and $\sim 1/4$
and $\sigma_1$ and $\sigma_2$, $\sim 0.37$ cm and $\sim 1.35$ cm. 
For distances larger than $\sim 2.1$ cm, the detected hits are tagged as background and 
a uniform probability density function is associated, $B(r) = \frac{b}{D}$, where
$D$ is an effective distance in the detector corresponding to $134$ cm and $b$ the background
fraction estimated as $77\%$.
The overall probability density function is therefore defined as:
\[
\mathcal P(r) = (1-b)\; S(r) + B(r)
\] 
The best value for the Cerenkov angle $\theta_c$ will result from the maximization
of a likelihood function $\mathcal{L}(\theta_c)$, built from the product of the event hits
probability weighted by the photon multiplicity (for $n_{p.e}/hit > 1$),
\vspace{-0.2cm}
\[
\mathcal L \mathnormal (\theta_c)=\prod_{i=1}^N \mc{P}^{n_i} \mn [r_i(\theta_c)]
\]
Figure~\ref{fig:betarec-vs-z} shows the velocity accuracy as function of the electric 
charge obtained for simulated data samples.
As expected, the larger is the number of radiated photons the better is the resolution 
with a systematic limit $\sim 2 \; 10^{-4}$ from the detector pixelization. 

\begin{figure}
\begin{center}
\noindent
\vspace{-0.6cm}
\includegraphics*[width=0.30\textwidth,angle=0,clip]{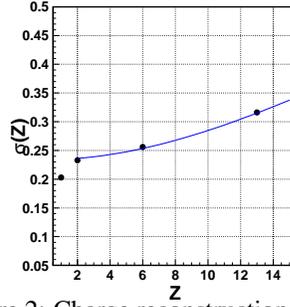}
\caption{Charge reconstruction accuracy. 
\label{fig:chargerec-vs-z}}
\end{center}
\vspace{-0.8cm}
\end{figure}

Photons are uniformly emitted along the particle path ($L$) and  
their number depends on the particle charge ($z$) 
and velocity ($\beta$) and on the refractive index of the medium ($n$),
\begin{equation}
\frac{dN_\gamma}{dE} \propto z^2 \; L \; \left( 1 - \frac{1}{\beta^2 \; n^2} \right)
\end{equation}
The fraction of photons in the ring pattern that are lost in every event 
depends on their topology (impact point and particle direction and velocity) 
and other factors. 
Therefore, charge reconstruction relies on the reconstructed \ck angle and on
measurements of the path length crossed by the particle,  
of the number of ring associated photoelectrons detected on the readout matrix 
and finally on the evaluation of the photon detection efficiency. 
The efficiency factors involve the radiator interactions, the photon
ring acceptance including the mirror reflectivity and the light guide and photomultiplier quantum
efficiencies. 
Therefore, applying the correction factors on an event-by-event basis,
the detected signal for singly charged particles ($N_0$) can be estimated 
and the charge of the incident particle is obtained according to:
\begin{equation}
Z^2=\frac{N_{p.e.}}{N_0}\propto
\frac{N_{p.e.}}{\varepsilon_{TOT}}\frac{1}{\Delta L}\frac{1}{\sin^2\theta_c}.
\end{equation}
Figure~\ref{fig:chargerec-vs-z} shows the charge estimated accuracy obtained from simulated
data samples for different nuclei. 

\vspace{-0.2cm}
\subsection{Detector monitoring during flight}
\vspace{-0.3cm}
The accuracy of the velocity and charge reconstruction provided by the RICH
relies on the precise knowledge of the detector parameters throughout the AMS
mission. For instance, the refractive index of the individual aerogel pieces 
must be known at the $10^{-4}$ level not to compromise the velocity measurement.
Similarly, the single PMT response must be known at the percent level not to
degrade the charge determination capabilities. Extensive characterization of
all optical elements was performed prior to and during detector assembly~\cite{bib:rich-Guia}. 
However, the launch and, in particular, the varying environmental conditions 
along the mission (e.g. temperatures ranging from -20$^\circ$C to 40$^\circ$C) 
may affect the optical properties of the detector parts. Therefore, specific
algorithms have been developed to monitor RICH optical parameters using flight
data. MC simulations show that the radiator refractive index will be monitored 
to the required precision level using 1-day equivalent electrons or high energy 
protons and the PMT gains will be determined with the adequate precision using
a 1-orbit equivalent (90 min.) proton data~\cite{bib:rich-Crespo}.
\vspace{-0.2cm}
\section{Physics prospects: D/p separation}
\vspace{-0.3cm}
The measurement of the deuteron/proton ratio provides information on cosmic-ray
production and propagation and is particularly challenging due to the low deuteron
abundance (D/p~$\sim$~1\%). 
The extremely accurate velocity measurement provided by the RICH 
($\Delta \beta / \beta \sim$ 10${}^{-3}$) is crucial to reduce the background level.
A full-scale simulation of the AMS detector
was used to evaluate the capabilities of AMS-02 for mass
separation. 

Pre-selection cuts using readings from different subdetectors of AMS-02 were 
applied to reduce the fraction of events badly reconstructed in momentum.
The set of events passing these cuts corresponds to an acceptance of
$\sim$~0.3 m${}^2$sr and $\sim$~0.2 m${}^2$sr, respectively for protons 
and deuterons.

The reconstruction of particle masses was then performed for events having
a signal in the RICH detector. A series of event selection cuts were introduced,
namely a minimum of 3 hits in the reconstructed Cerenkov ring, an upper limit
on the number of noisy hits and compatibility between two independent
velocity measurements.

\begin{figure}[htb]
\center
\includegraphics*[width=0.5\textwidth,angle=0,clip]{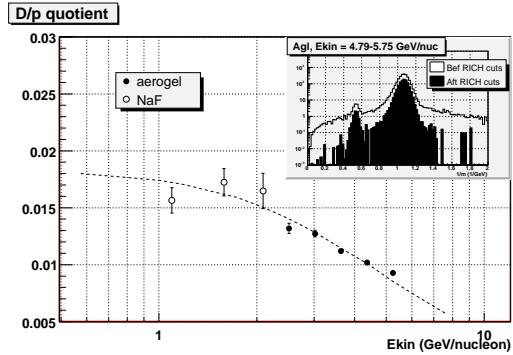}
\vspace{-0.6cm}
\caption{Reconstructed deuteron/proton ratio for the simulated sample.
The dashed line shows the simulated ratio.
\label{dpratio}}
\vspace{-0.2cm}
\end{figure}

\begin{figure}[htb]
\center
\includegraphics*[width=0.5\textwidth,angle=0,clip]{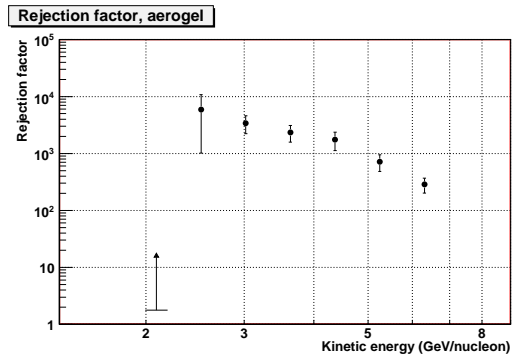}
\vspace{-0.6cm}
\caption{Rejection factor for D/p separation in aerogel events.
\label{rejfac}}
\vspace{-0.4cm}
\end{figure}

Results show that mass separation of particles with $Z=$~1 is feasible even
if one species is orders of magnitude more abundant than the other. It is
possible to obtain good estimates for the D/p ratio in the few
GeV region (Fig.~\ref{dpratio}).
D/p separation is possible up to $E_{kin} \sim$~8 GeV/nucleon. 
In the optimal region immediately above the aerogel radiation threshold
($E_{kin} =$~2.1~$-$~4 GeV/nucleon) rejection factors in
the 10${}^3$~$-$~10${}^4$ region were attained (Fig.~\ref{rejfac}). 
The best relative mass resolutions for protons and deuterons
are $\sim$~2\% for both radiators in the regions above their respective
thresholds.
After all cuts, an acceptance of $\sim$~0.06 m${}^2$sr and 
$\sim$~0.04 m${}^2$sr was obtained, respectively for protons and deuterons 
for kinetic energies above 3 GeV/n~\cite{bib:rich-dp-ecrs06}. 

\vspace{-0.4cm}
\section{Conclusions}
\vspace{-0.3cm}
The AMS experiment includes a Ring Imaging Cerenkov detector.
Algorithms for charge and velocity reconstruction were developed, confirming
the design capabilities of the instrument: $\sigma_{\beta} / \beta \sim 0.1$\% 
for singly charged particles and $\sigma_z \sim 0.2$.
In order to achieve such accuracy, an extensive characterization on laboratory of all 
optical elements was done and monitoring of the aerogel refractive index and
photomultiplier gain will be performed during flight.
Analysis based on simulated data samples, show that despite the large background,  
separation of deuterons from protons is possible up to kinetic energies around 8 GeV/n.


\end{document}